# Spin chemistry of *sp²* nanocarbons

Elena F. Sheka

Peoples' Friendship University of Russia (RUDN University), 117198 Moscow, Russia

sheka@icp.ac.ru

**Abstract**

Suggested in the paper is the consideration of stable radicals of *sp²* nanocarbons from the standpoint of spin-delocalized topochemistry. Characterized in terms of the total and atomically-partitioned number of effectively unpaired electrons as well as of the distribution of the latter over carbon atoms and described by selectively determined barriers of different reactions exhibiting topological essence of intermolecular interaction, *sp²* nanocarbons reveal a peculiar topokinetics that lays the foundation of the stability of their radical properties.

1. Introduction

Spin chemistry is, generally speaking, the chemistry of open-shell molecules. The total pool of molecular spins consists of $N_\alpha$ and $N_\beta$ spins, oriented up and down, respectively, and the difference $SpM = N_\alpha - N_\beta$ determines spin multiplicity of the molecule ground state. Predominant majority of α and β spins form pairs, both components of which are located in the same space subordinating to the Pauli principle and thus forming closed-shell spin-orbitals. When a part of α and β spins is located in different space, the formed spin-orbitals are open-shell, the same are the relevant molecules. Certainly, when $SpM \neq 0$, the open-shell character of molecules is obvious. However, when $SpM = 0$, the problem arises to classify molecules as either closed- shell or open-shell ones. This problem of the singlet-ground-state molecules is the main topic of the current study.

The history of quantum-chemical consideration of open-shell molecules is quite long. It originates in 65 years ago from the paper of Pople and Nesbett [1], who suggested *Unrestricted Hartree-Fock* (UHF) approximation to describe electronic properties of radicals, which has remained conceptually the best until now. First applications of the approach to singlet open-shell species revealed a characteristic UHF feature concerning spin contamination (SC) of the spin multiplicity of the ground states in the term of the total squared spin profit $\Delta \hat{S}^2 [= \hat{S}^2_{UHF} - \hat{S}^2_{exact}]$ that is the most pronounced in the singlet case due to $\hat{S}^2_{exact}=0$. The SC undoubtedly evidences the breaking of spin symmetry of the electronic system. Prominent Löwdin's papers [2, 3] allowed establishing a direct connection between SC and/or spin symmetry breaking with electron correlation that was not taken into account in the UHF framework to full extent. Since this point, the development of the quantum chemistry of open-shell molecules was divided into two branches. The first stream (CI - branch herein) was based on the wish to restore broken spin symmetry substituting two-determinant UHF computational algorithm by multi-determinant ones that provide accounting of configurational interaction (CI) [3]. Changings concerned only wave functions that became more and more complicated while Hamiltonian of the systems remained unchanged. Many computational schemes has been suggested on this way. A most recent review [4], presented by Chattopadhyay's team, gives quite complete description of the CI-branch modern state of art and supplies readers with many valuable references.

The second UHF branch was based on the acception of broken spin symmetry in open-shell molecules as a reality and of SC as a characteristic mark of systems of this type. The approach was originated by Takatsuka, Fueno, and Yamaguchi [5] and was supported by Davidson and his coworkers [6]. Forty-year development of the approach led to its formulation in terms of emergent phenomena in many-electron systems [7]. Conceptually, this view of electronic systems is as follows. When many-electron system is spin symmetric the interaction between electrons adiabatically reduces to the sum of solutions for individual electrons. This is typical for closed-shell electron systems and fully realized in restricted Hartree-Fock (RHF) approximation. The same is apparently looked for open-shell systems in the high-order CI-schemes. When the symmetry is broken, the solution for interacting electrons does not obey the laws; as a result, there are new residual (so-called *emergent*) phenomena. Thus, the key question to answer is whether these residual phenomena can be regarded as compatible with physical reality. The answer divides quantum chemists into reductionists and emergentists, saying NO and YES, respectively. This view on the situation with open-shell molecules was introduced to scientific community by the Nobel Prize winner in Physics 1977 P.W. Anderson in his famous article ``More is different'' [8] just based on UHF solutions. The idea was supported and continued by one more Nobel Prize laureate in Physics 1998 R.B. Laughlin [9], and opened the door to a realm of unexpected phenomena [10-13]. As turned out, in physics emergent phenomena resulted from symmetry breaking are quite numerous and well known. Enough to mention emergent quasiparticles such as phonons, excitons, solitons, polarons, anyons, and so forth as well as physical phenomena such as topological conductivity, magnetism, and superconductivity. It was proposed in Ref. [7] to extend the emergent concept on chemistry as well addressing it to the consideration of open-shell molecules in terms of UHF emergents. Such an approach not only supplies a computationist with classifying markers which make the treatment well ordered and conceptually self-consistent, but has one more advantage over CI-branch, which is provided with high efficiency of the UHF algorithm, particularly implemented in semi-empirical versions allowing computational treatment of large molecules.

On practice, in computational spin chemistry of open-shell molecules there has been a paradoxical situation. Quantum chemists are mostly committed to the reductionist concept of the CI- branch. However, the practical application of advanced and most accurate CI-methods is still limited to quite small molecular systems while the real chemistry requires the consideration of large electronic structures containing above hundred atoms. Meeting these requirements, a lot of effort was made to develop approximate but time-consuming techniques. At the same time, it so happened historically that the UHF approximation was publicly repeatedly declared conceptually opposite to the mainstream CI ones. Its inherent emergents were heralded as fundamental errors [14, 15] or even unphysical [16]. Expectedly, the approach has been excluded by the main stream from the proper techniques and the empty place was quickly occupied by approximate methods, such as, but a few, density matrix renormalization group (DMRG) [17], restricted active space spin–flip (RAS-SF) [18], and various versions of DFT. If the first two methods could still be considered as some advance towards exact CI methods, then the DFT threw computationists far back into the domain of one-determinant methods. Notwithstanding, the quantum chemical community not only did not resist this seizure as was in the case of UHF, but encouraged and approved it in every possible way. It did not matter that the DFT in all its aspects not only does not approach CI-methods, but significantly inferiors to the UHF approximation. If for closed-shell molecules, this drawback could be compensated by the special empirically based adaptation of functionals, which makes the technique fully empirical, it turned out that DFT wasn't cut out for open-shell molecules at all (see a detailed discussion of the UDFT problems related to the case [19-21]). Similarly to UHF concept, DFT refuses to spin symmetry when transforming into unrestricted (UDFT) version that is symmetry broken. In addition, UDFT algorithm operates with spins in non-direct and much more complicated way and touches on not only wavefunctions, but Hamiltonian as well thus making the restoration of spin symmetry absolutely impossible [8]. In all the cases of comparative studies, which used UHF, some of CI-approaches, and UDFT (see a brief review of the issue in [7] as well as the most recent publication [22]), UDFT demonstrated the worth results with respect to the former two techniques which, in contrast, showed practically identical results. Despite these obvious and well known unfavorable circumstances, over the past two decades, DFT spin chemistry has turned into massive virtual computational chemistry of open-shell molecules, which has flooded hundreds and thousands of scientific publications. Not only UDFT, but widely used restricted DFT as well can be met on these pages. To date, the capabilities, availability and effectiveness of DFT methods have been compared with experimental synthetic and analytical procedures, as a result of which they have become an integral part of practical chemistry.

It was necessary to happen that this period coincided with the time when special molecules appeared on the stage of modern chemistry - $sp^2$ nanocarbons, including fullerenes, carbon nanotubes and graphene leaf fragments. All the molecules contain even number of electrons. Spin multiplicity of the ground state of all molecules should begin with singlet. A whole army of quantum chemists, equipped with the most modern versions of the DFT method, which are in the public domain, rushed to storm new peaks. Quasi CI techniques DMRG and RAS SF, efficient enough to treat a large class of open-shell molecules related to polynuclear aromatic hydrocarbons (PAH) (see review [23]), turned out to be inapplicable to $sp^2$ nanocarbons. Since all these molecules are unique both in their properties and in the handling methods, experimental studies have been noticeably lengthened and became more complicated, which led to the dominant contribution of DFT-based calculations to this field of chemistry, dividing it into real and virtual. Therewith, virtual fullerenics and graphenics confidently explains and predicts everything: new properties, new materials, and, most importantly, new applications. But real chemistry lags behind and does not reveal these new properties, materials and applications. This discrepancy was especially noticeable during the fulfillment of one of the programs with the largest budget for material science known to date that is the international program "Graphene Flagship" [24]. Serious complications met at fulfilling given promises first led to the appearance of the terms of 'good' and 'bad' (high-performance and low- performance) graphene [25], and then six years later the program founders explained the failure by the dishonesty of graphene material manufacturers when producing good graphene [26]. According to our standpoint, the true reason of the problem concerns spin chemistry and is due to the incorrect concept of graphene as a technological material, which was imposed by the unlimited dictate of virtual DFT graphenics. The true properties of graphene turned out to be undisclosed while presentations of its chemical and physical properties was erroneous. In practice, graphene really behaves as a radicalized object, the spin essence of which the DFT was unable to notice. Even the latest publication, concerning a seemingly coordinated attitude to the radical properties of small graphene fragments [27], they are considered with an indispensable eye on the DFT. At the same time, remaining in a shadow and parallel developed UHF concept of computational graphenics has acquired new evidence of the legitimacy of the approach and the results obtained [28-36]. The semi-empirical UHF method has proven to be very effective and allowing to get valuable information for large open-shell molecules. The efficiency of the UHF calculations is high so that calculations of, say, fullerene $C_{60}$ or $C_{70}$ take not more than an hour on an advance personal computer while systems of 200-400 atoms can be considered for one-three days. This allows performing not only single computations, but computational experiments on numerous objects to prove the words of one of the Nobel Prize laureates in chemistry R. Hoffman: "It goes without saying that theory is really of value when it is used to perform numerical experiments that capture a trend. Not numbers, but a trend" [37]. The current paper is devoted to trends in spin chemistry of $sp^2$ nanocarbons revealed by UHF emergent and supported by numerous experimental evidences. The paper presents the issue from the standpoint of the user of computational quantum-chemistry, based on the experience gained during about thirty years.

## 2. Grounds of UHF computational spin chemistry

### 2.1. General remarks

Naturally, the combination of words 'spin chemistry' carries an understanding of the special role of electron spins in the ongoing chemical transformations. To make this process visible from the viewpoint of the UHF-spin chemistry, we shall divide it into four conditional stages that present the main issues of practical chemistry.
- Spin traits of the reactants that enter into reactions;
- Spin kind of the intermolecular interaction that controls such reactions;
- Spin nature of the reaction final products;
- Post-reaction existence of spin chemistry products.

Before to start, we have to clearly realize which features of the above issues will be described in terms of UHF emergents. Among many others, we will mainly concentrate on those that are related to the equilibrium ground state leaving aside continuous symmetry problems caused by spin symmetry breaking [38] that drastically change the appearance of optical electron [34] and vibrational [39] spectra of open-shell molecules. Concerning the ground state, the following features are of interest:

- Ascertainment of the radical status of an open-shell molecule;
- Evaluation of the molecule chemical activity as a whole;
- Establishment of the spin density delocalization over the molecule atoms;
- Determination of the chemical activity over molecule atoms thus presenting its 'chemical portrait';
- Detection of the spin-chemical topology of the molecules.

The following quantities calculated during the computational experiment are the source of the required information. At the first place there are the molecule total energies $E_{sg}^{RHF}$, $E_{sg}^{UHF}$, $E_{tr}^{UHF}$, and two differential energies $\Delta E_{sg}^{RU}$ (=$E_{sg}^{RHF}$ - $E_{sg}^{UHF}$) and $\Delta E_{ST}^{UHF}$ (=$E_{tr}^{UHF}$ - $E_{sg}^{UHF}$) (subscripts $sg$, $tr$, and $ST$ mark singlet and triplet states and singlet-triplet energy gap, respectively). The second place is occupied by emergent spin characteristics $\Delta \hat{S}^2$ (= $\hat{S}_{UHF}^2$ - $\hat{S}_{exact}^2$), $SpD_{tot}$ and $SpD_A$ as well as $N_D$ and $N_{DA}$ ($SpD_{tot}$ and $N_D$ describe total spin density and total number of effectively unpaired electrons [5, 6] while subscript $A$ matches these values related to atom A). A complete set of {$sp^2$C-C bonds} closes the necessary suit of source data. In what follows the description of the above issues will be carried out on the example of spin chemistry of a representative set of $sp^2$ nanocarbon molecules, including fullerene C$_{60}$, a fragment of a single-walled carbon nanotubes (SWCNT), and honeycomb-structure compositions presenting a set of graphene molecules.

## 2.2. Common background for $sp^2$ nanocarbons features

Despite the fact that {$sp^2$C-C bonds} pool ends the above list of initial data, it is the network of these bonds, that is the general structural motive of substances as well as the main reason of their features. The bonds are quite labile and may change the length in the range of 1.326 -2.158 Å (see Chapter 2 of the monograph [36] and [40]). On this way, they transform from covalent non-radical to fully radicalized ones. The radicalization extent depends on the current bond length and starts from zero at the critical bond length $R_{crit}$, which slightly depends on the bond surrounding and changes from 1.395 to 1.408 Å when going from ethylene to hexamethylbenzene, and drastically grows when the length exceeds $R_{crit}$. The transformation of the bonds from covalent to radicalized ones leads to the transition of closed-shell electronic structure to open-shell one. Certainly, we begin the consideration of the chosen nanocarbon molecules from the description of {$sp^2$C-C bonds} distribution.

It is obvious that in the case of extended bond networks, not only individual bond lengths, but also a composition of the bonds in pool is significant. The latter particularity is clearly seen in Fig. 1 where the bond distribution of three examples are presented. Inserts in the figure represent equilibrium structures of fullerene $C_{60}$, a fragment of (4, 4) SWCNT and right-angle fragment of a flat honeycomb structure with 5 benzenoid units along both zigzag and armchair edges, (5, 5)NGr molecule herein. As seen in the figure, in all the cases the bond lengths cover large region of ~0.3 Å in width and the bonds with length over $R_{crit}$ make up more than half of the pool. Evidently, the presence of these bonds lays the foundation of the open-shell character of the molecules. However, the generality of the molecules in question ends there. The distribution plottings are quite different and deserve a particular consideration.

*Fullerene $C_{60}$.* Widely accepted presentation about the molecule of local symmetry $I_h$ is supported by RHF calculations only. The exact symmetry of the molecule in the UHF approach is $C_i$ [38] and the symmetry difference is clearly evident in Fig. 1a. It should be noted that a division of bonds into short and long ones remains in both cases and the difference concerns these bonds dispersion that is 1.385 ± 0.0002 Å and 1,463 ± 0.003 Å for the bond pairs in the RHF solution and 1.391 ± 0.032 Å and 1,464 ± 0.013 Å in the UHF case. Obviously, the dispersion changes are not too drastic, due to which the symmetry changing in fullerene $C_{60}$ presents a perfect example of continuous symmetry when the lower symmetry, in fact, remains the high to a large extent [42]. In the current case, $C_i$ symmetry of $C_{60}$ includes 95% of $I_h$ [38]. Nevertheless, the deviation from the $I_h$ is definitely revealed experimentally by the appearance of forbidden transitions in optical [38] and vibrational ('silent modes') [39] spectra of the molecule.

Important to note that the difference in the RHF and/or UHF description of the molecule {$sp^2$C-C bonds} pool concerns not only the change of equilibrium symmetry, but the reaction of the pool to any effect on the molecule. In the case of closed-shell approximation, each individual addition is local and does not disturb the distribution of the remaining bonds. In contrast, UHF approach reveals a considerable rearrangement of the whole bond pool at each case of the intrusion thus demonstrating a collective character of the molecule valence electron structure (see detailed discussion of the feature on many examples in [43]).

*(4, 4)SWCNT*. The two-length bond structure is characteristic for a series of PAHs and is commonly considered as a characteristic mark of the molecule aromaticity. Accordingly, until now, sometime a referring to fullerene $C_{60}$ as aromatic specie can still be met in literature. However, the format of two-length bond structure strongly violates in the case of $C_{70}$ and fully disappears in CNTs and graphene molecules. Figure 1b exhibits the picture of {$sp^2$C-C bonds} distribution of the (4, 4)SWCNT fragment determined in two approximations [44]. In contrast to $C_{60}$, closed-shell model reveals 10 groups of the bonds with small dispersion within each group and quite large changing between the groups. Open-shell approximation leads to a remarkable ordering of the tube bond structure decreasing the group number practically by half. Particularly noticeable is the group in the plotting middle covering a continuous changing the length from 1.436 to 1.431 Å which is related to the tube side wall. Actually, when the fragment length increases, this group respectively grows as well. Both distributions in Fig. 1b also reveal short-length groups related to the tube cap and long-length ones attributed to the tube open end. When the cap is substituted with open end, the short-length bonds disappear from the plotting. This cap-sidewall-end kind of the {$sp^2$C-C bonds} plotting remains when tubes diameter increases as well as when (n, n) tubes are substituted with (n, m) ones [44]. The reaction of the {$sp^2$C-C bonds} on chemical addition depends on the locality of the latter. In the sidewall region, the response is quite local while a considerable disturbance is characteristic for caps and open ends.

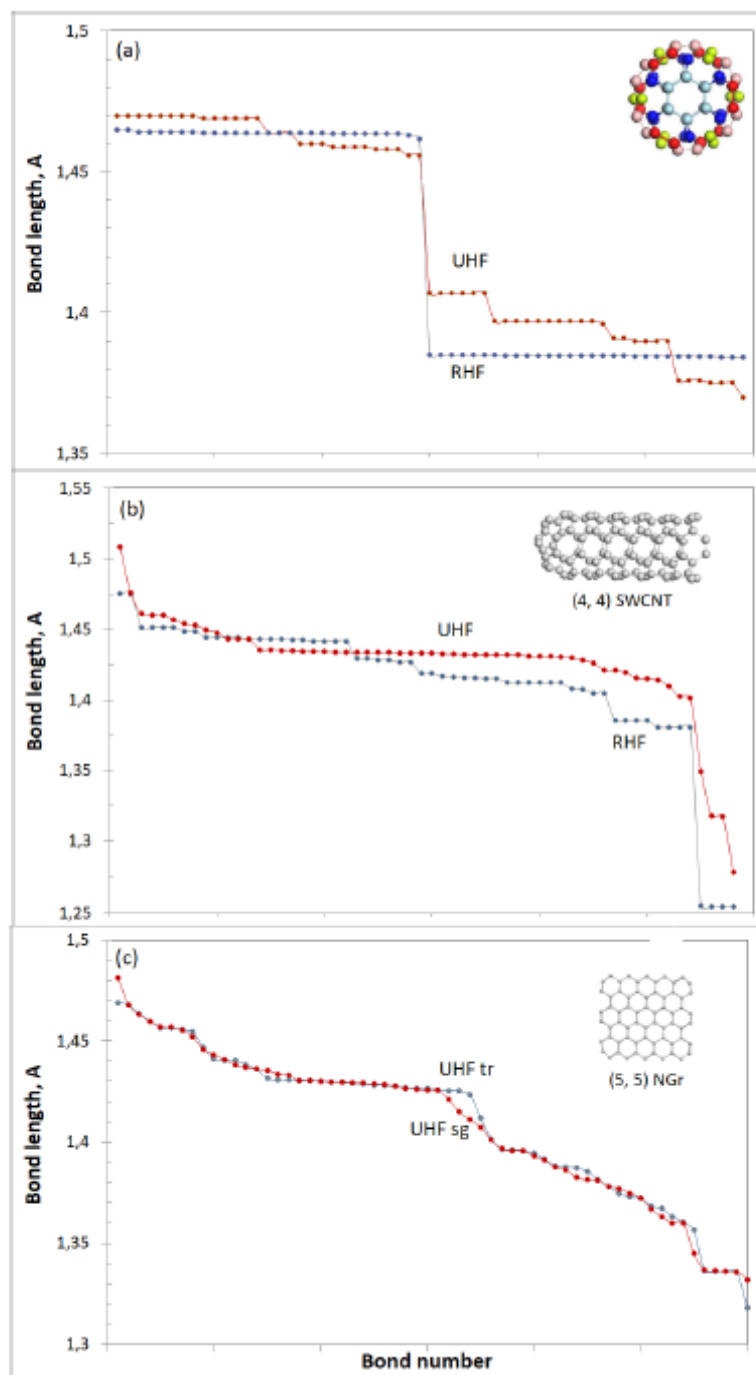

**Figure 1**. Z→A distributions of $sp^2$C-C bonds in fullerene $C_{60}$ (a), a fragment of (4, 4)SWCNT (b) and nanographene molecule (5, 5)NGr (c). calculations are performed using CLUSTER-Z1 software implementing AM1 semi-empirical HF approximation [41]. Closed-shell version of the study is presented by the RHF solutions.

*(5, 5)NGr molecule*. A comparative pattern of {$sp^2$C-C bonds} of the closed- and open-shell compositions of the molecule is similar to that one discussed above for the tube with the only difference related to the absence of characteristic 'sidewall' bonds. Extending the molecule size leads to increasing the groups number while smoothing the difference between the groups, which results in a continuous decreasing of bond lengths from ~1.49 Å to 1.33 Å in the relevant Z→A plottings. One more difference is characteristic for graphene molecules. It concerns the response of the {$sp^2$C-C bonds} pool on the chemical addition, which drastically changes the pool distribution at each action (see a detailed discussion of the feature on various examples in [36]).

Besides the features discussed, Fig. 1c draws attention to an exclusive particularity of the molecule. It concerns extremely small singlet-triplet gap $E_{ST}$ which in the case of (5, 5)NGr, constitutes -6.096 kcal/mol. The triplet state is slightly lower than singlet, which means the two states are strongly mixed. Applying UHF approach to both states separately, we obtain quite similar {$sp^2$C-C bonds} plottings, which is clearly seen in the figure, thus justifying that at the plotting level the two states cannot been distinguished.

Concluding the description of {$sp^2$C-C bonds} pools, we can draw the following conclusions.
1. More than half of $sp^2$C-C bonds of $sp^2$ nanocarbons, including fullerenes, carbon nanotubes and graphene molecules, are longer than critical interatomic distance $R_{crit}$, exceeding over which leads to the bond radicalization. The feature lays the foundation of the open-shell character of the species electron systems and radical character of the molecules
2. If two-length bond composition is characteristic for the {$sp^2$C-C bonds} pool of fullerene C$_{60}$, many-length one is typical for more extended {$sp^2$C-C bonds} networks starting from fullerene C$_{70}$ and involving CNTs and graphene molecules.
3. Application of UHF approach allows disclosing a collective response of {$sp^2$C-C bonds} pool to each act of any chemical addition to the relevant species in all the cases thus revealing delocalization of the bond distribution disturbance.

### 3. Spin traits of open-shell molecules

#### 3.1. Fullerene C$_{60}$

Having dealt with the {$sp^2$C-C bonds} pool of the chosen molecules, we now turn to data concerning their energies and spin emergents. The collection of the corresponding source data is presented in Table 1. The data evidence convincingly that the considered molecules are of the open-shell type and their chemistry is spin by nature.

**Table 1.** Energies (kcal/mol) and emergent spin characteristics of molecules in the ground state

| Molecules | $E_{sg}^{RHF}$ | $E_{sg}^{UHF}$ | $\Delta E_{sg}$ | $E_{tr}^{UHF}$ | $\Delta E_{ST}^{UHF}$ | $\Delta \hat{S}^2$ | $SpD_{tot}$ | $N_D$ |
|---|---|---|---|---|---|---|---|---|
| C$_{60}$ | 970.180 | 955.362 | 14.818 | 982.917 | 27.555 | 4.92 | 0 | 9.84 |
| (4,4)SWCNT | 1819.509 | 1648.338 | 171.171 | 1627.308 | -21.030 | 19.758 sg<br>20.325 tr | 0 | 39.515 sg<br>38.651 tr |
| (5,5)NGr | 1802.381 | 1454.492 | 347.889 | 1448.396 | -6.096 | 16.930 sg<br>18.121 tr | 0 | 33.860 sg<br>34.242 tr |

The first acquaintance with UHF spin chemistry is best to start with $C_{60}$ fullerene, which has all characteristic emergents of open-shell molecules, on the one hand, and is a real well known molecule, on the other, which allows addressing experimental data to confirm computationally predicted emergent features. As seen in Table 1, UHF approach remarkably lowers the energy of singlet state, which is much lower than triplet is and constitutes itself as a ground state. The singlet-triplet gap accounts for ~3% of the total energy and is enough for the spin contamination of the ground state to be clearly fixed. Due to singlet character of the ground state, the total number of effectively unpaired electrons $N_D$ equals to doubled spin contamination $\Delta \hat{S}_{sg}^2$. This number is distributed over the molecule atoms in terms of partial numbers $N_{DA}$, plotting of which over the molecule atoms is shown in Fig. 2a in two manners, namely, as histogram presenting Z→A distribution and dotted curve related to the atom numbers in the output file. The first plotting allows distinguishing clearly seen six groups of atoms with 12 atoms of the same $N_{DA}$ value per each group of the first four groups and with 6 atoms of groups 5 and 6. Atoms of the first group form two hexagons (see configurations of slight blue atoms in inserts in Figs. 1a and 2a), atoms of groups 2, 3 and 4 are joint by pairs (red, rose and green) while dark blue atoms are singles [45]. Marked by different colors, the atoms form a multicolor image of the molecule shown as insert in Fig. 2a. Plotted in Fig. 2b presents A→Z spin density distribution that clearly demonstrates antiferromagnetic delocalization of the density over the molecule atoms providing the $SpD_{total}$ exact zeroing.

Since $N_{DA}$ presents atomic chemical susceptibility (ACS) [46], the insert image presents a color view of the chemical reactivity distributed over the molecule atoms from the most reactive (light blue) to the most inactive (dark blue) thus allowing to draw the fundamentals of spin chemistry fullerene $C_{60}$. The first element of the set states that the molecule is chemically active and its molecular chemical susceptibility (MCS) $N_D$ constitutes 9.87 e. According to the second element, the molecule is multi-target chemical object since all its atoms can participate in the chemical reaction with a fractional contribution $N_{DA}$. Following the third element, the molecule enters any chemical action by atoms of the highest chemical activity described by the largest $N_{DA}$ value. Since $N_{DA}$ distribution over the molecule is spatially peculiar and changes at each reaction step, a chemical reaction, involving fullerene $C_{60}$, acquires topological features depending on the chemical counteragent, which is the fourth element of the $C_{60}$ spin chemistry. The fifth element concerns the polyderivative character of the reaction between any reagent and fullerene. The sixth one determines the derivative order related to the reaction completion which is due to continuous decreasing of the molecular chemical susceptibility $N_D$ in the course of the reaction until it is worked out up to nil. Evidently, the reaction may be stopped before $N_D$ reaching nil due to sterical obstacles.

### 3.2. (4 4)SWCNT fragment

As turned out, the fundamentals discussed in the previous section govern generally the spin chemistry of not only fullerene $C_{60}$ as well as higher fullerenes [43], but other *sp²* nanocarbons such as CNTs and graphene molecules. As seen in Table 1, the UHF approximation eliminates the artifact when RHF singlet state is located much above the triplet one and displaces the singlet down. The energy gap $E_{ST}$ is small and makes up 1.3% of the energy of the singlet state. For the current case, the gap is negative indicating that formally triplet should be attributed to the ground state. Notwithstanding, the sign of the gap is not steady and changes when either length or diameter of the tube increases. The only steady point is that the gap is small so that singlet and triplet states are mixed. It should be noted that the spin contaminations of both states, $\Delta\hat{S}_{sg}^2$ and $\Delta\hat{S}_{tr}^2$, as well as the total numbers of effectively unpaired electrons, $N_{Dsg}$ and $N_{Dtr}$, are considerable and practically the same. The spin-triplet nearly-degeneracy, which is typical for polyradicoloids with double C-C bonds [22, 23], greatly complicates the description of open-shell electronic states [4, 20, 21] and a proper computational technique has been still absent. Therefore, we will have to limit ourselves to the available capabilities and hold a discussion of the question posed in terms of the UHF consideration of the singlet state.

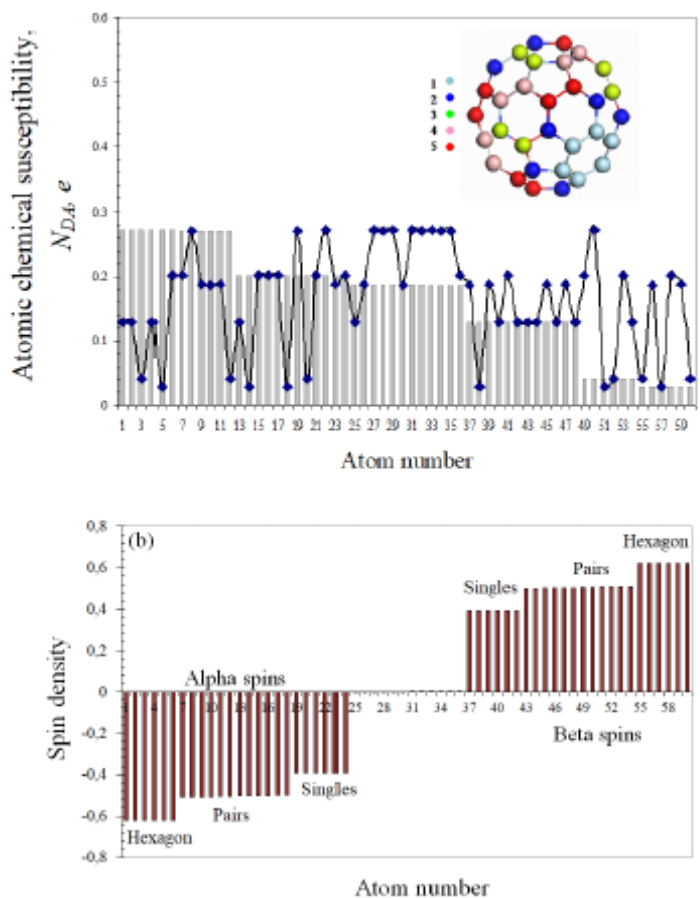

**Figure 2**. (a). Atomic chemical susceptibility $N_{DA}$ of fullerene $C_{60}$, distributed over the molecule atoms according to either their numeration in the output file (curve with rhombs) or in the Z→A manner (histogram). Different colors in the insert on right distinguish six atomic groups shown by the histogram. (b). Spin density distribution. Adapted from [45].

The description of the (4, 4)SWCNT molecule as an open-shell system is much the same as for fullerene $C_{60}$. Plotting in Fig. 3 presents the ACS $N_{DA}$ distribution over the molecule atoms. The distribution convincingly tell us that i) the tube is a multi-target object; ii) the main reaction ability is concentrated on open-end atoms and these atoms enter any reaction always first; iii) less active atoms are located in the cap zone and still less active atoms are distributed along sidewall. Different from the fullerene case is the extremely high ACS values for open-end atoms. The feature is connected with the fact that ACS of cap and sidewall atoms is determined by the {$sp^2$C-C bonds} pool, due to which their ACS is comparable with that one of fullerene atoms, while ACS of open-end atoms are additionally provided by dangling bonds (DBs). Actually, as seen in the figure, saturation of these DBs by monoatomic hydrogens drastically inhibits the end atoms reactivity equalizing it with the sidewall one. The discussed fundamentals govern all general features of the CNT spin chemistry, which are widely confirmed experimentally [48].

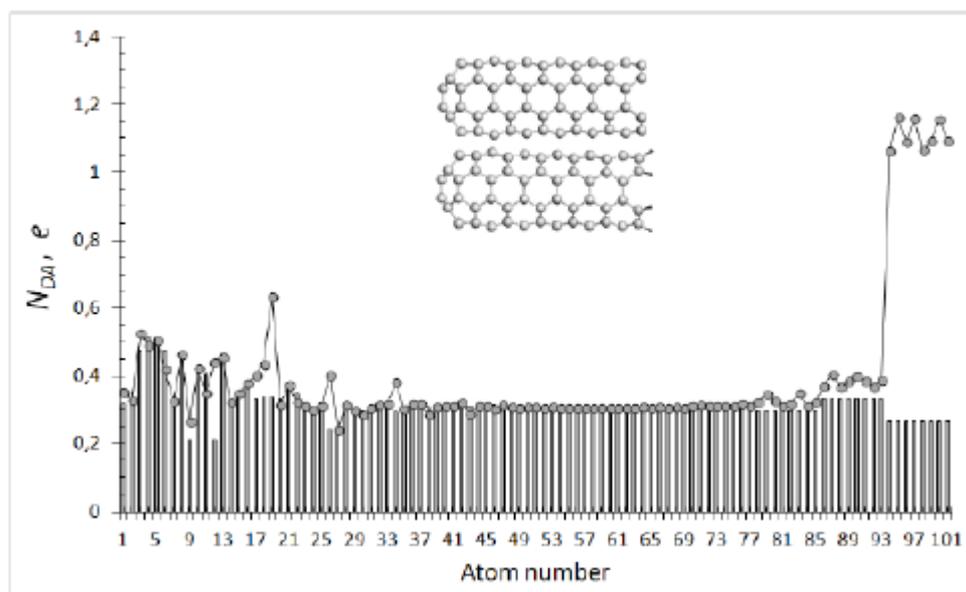

**Figure 3.** Standard ACS $N_{DA}$ image maps over atoms of (4, 4) Single walled carbon nanotubes shown in insert with empty (curve with dots) and hydrogen-terminated (histogram) ends. The numbering follows a movement along the tube atoms from the cap to end.

### 3.3. (5, 5)NGr molecule

As seen in Table 1, everything that was said earlier with respect to the singlet-triplet degeneracy of the CNT states can be fully addressed to graphene molecules. In the latter case, the degeneracy becomes even more pronounced since the $E_{ST}$ gap still more decreases and accounts for 0.4% of the total energy. Expectedly, the sign of $E_{ST}$ can easily alternate and, in fact, this is what happens when size, shape and chemical modification in the area of edge atoms change. Leaving for the future a correct theoretical consideration of open-shell systems with nearly-degenerate states, we restrict ourselves below to the discussion of graphene molecules in the framework of the UHF consideration of one of the degenerate states. As occurred in the current study, the consideration of either singlet or triplet state leads to practically the same results that might be caused by the close equality of $\Delta \hat{S}^2$ and $N_D$ data for both states.

A considerable value of MCS $N_D$ evidences about high chemical activity of the (5, 5)NGr molecule as a whole. Distributed over the molecule atoms, it is presented as ACS $N_{DA}$ histogram plotting in Fig. 4. The spin density distribution [7] covers all molecule atoms as well and its plotting is qualitatively identical to that of fullerene $C_{60}$ in Fig. 2b with $SpD_{total}$ equal zero. As for $N_{DA}$ plotting seen in the figure, edge atoms dominate due to large contribution providing DBs. As in the case of CNT, considered in the previous section, their chemical reactivity can be inhibited by monoatomic termination of edge-atom DBs by hydrogens. In this case, the molecule becomes one of known polyradicaloids belonging to *peri*-anthracenes [23] and preserves its radical essence. Edge atoms mark zone of the highest reactivity while that one in basal plane is mainly determined by the {$sp^2$C-C bonds} pool and is at the level of fullerene and CNT sidewall atoms. The molecule as a whole is evidently multi-target prone to polyderivatization. Each step of polyderivatization is accompanied with the redistribution of remaining $sp^2$C-C bonds and, consequently, of $N_{DA}$ values thus evidencing a collective character of the electron system of the molecule. A drastic difference in chemical reactivity of edge and basal-plane atoms as well as a sharp anisotropy of the spatial structure in the direction normal to the basal plane provide extreme conditions for topological spin chemistry [48]. Taken as a whole, the described chemical behavior of graphene molecules is observed experimentally. Readers can find a number of examples in Ref. [36].

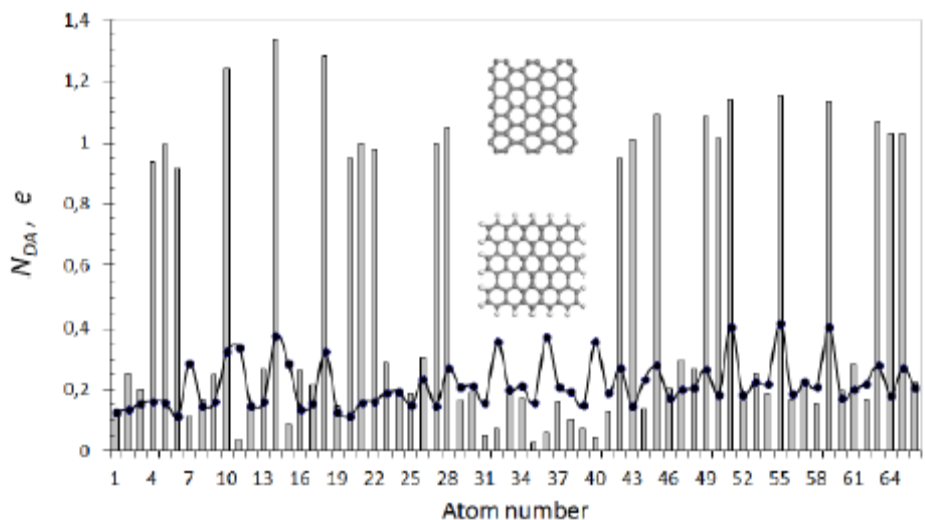

**Figure 4.** Standard ACS $N_{DA}$ image maps over atoms of (5, 5) NGr molecule with empty (histogram) and hydrogen-terminated (curve with dots) edges. Insert: equilibrated structures of the parent (5, 5)NGr molecule and its monohydrogenated homologue *peri*-anthracene.

All the peculiarities of molecules discussed above are rooted in the spin symmetry breaking. We share the view that the action is resulted from the quantum transition [9] that makes the UHF approach a good basis for the description of ground state, experimental justifications of which was discussed in [7] and has additional confirmation just recently [22]. Characterized in terms of UHF emergent, spin traits of the discussed molecules include the following issues. At atomic level, the latter concern a large set of local spins that lay the foundation of the over-atom delocalization of both spin density and chemical activity. At molecular level, high molecular chemical susceptibility, evidencing radical character of the molecules, and topological character of chemical reactions are the main issues. At energy level, small energy gap $\Delta E_{ST}^{UHF}$ of alternate sign draws attention to an exceptional feature concerning singlet-triplet nearly-degeneracy. The issue causes the necessary inclusion of the development of the quantum theory of open-shell electronic systems with pronounced singlet-triplet degeneracy in the agenda of modern quantum chemistry.

## 4. Electron spins in intermolecular interaction

When one of the partners of chemical reaction is of the open-shell type, the intermolecular interaction (IMI) between partners considered in the UHF approximation takes spins into account directly, operating with two determinants related to $\alpha$ and $\beta$ spin orbitals. The participation of spins in the formation of this interaction is hidden from the outside view and the correctness of their accounting can be judged only by results concerning, say, final products of the reaction. This aspect will be discussed in details in the next section. Herein we shall consider a particular case when spins are evidently involved in the IMI. The matter is that $sp^2$ nanocarbons molecules are not only open-shell species, but are characterized by exclusive donor-acceptor records, allowing them to be both donors and acceptors simultaneously. The issue is of particular importance when concerning IMI [43, 49]. Combining both open-shell character and donor-acceptor, IMI leads sometimes to an exciting result. Photodynamic (PD), effect directly concerning IMI of fullerene $C_{60}$ and molecular oxygen [50], can be the best example. The effect is widely used in the medicinal chemistry and consists in the transformation of triplet molecular oxygen into singlet one under photoexcitation. The presence of fullerene $C_{60}$ is therewith mandatory. Turning off the light returns the system to its previous state with inactive oxygen.

Many spears have been broken in attempts to explain the observed effect without taking into account the open-shell character of the fullerene (see [51, 52] and references therein). Until now, the mechanism of the effect has been hidden behind a slogan 'triplet state photochemical mechanism' that implies the excitation transfers over a chain of molecules according to a widely accepted scheme [52, 53]

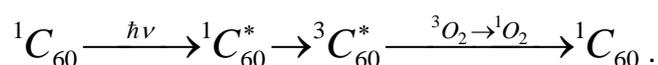

$$^1C_{60} \xrightarrow{\hbar\nu} {}^1C_{60}^* \to {}^3C_{60}^* \xrightarrow{{}^3O_2 \to {}^1O_2} {}^1C_{60}.$$

The scheme implies the energy transfer from the singlet photoexcited fullerene to the triplet one that further transfers the energy to convenient triplet oxygen thus transforming the latter into active singlet oxygen. The first two stages of this 'single-fullerene-molecule' mechanism are quite evident while the third one, the most important for the final output, is obscure in spite of a number of speculations available [53]. Obviously, this stage efficacy depends on the strength of the IMI between fullerene and oxygen molecules. Numerous quantum chemical calculations show that pairwise interaction in the dyad $[C_{60} + O_2]$ in both singlet and triplet state is practically absent. The UHF computations [54] fully support the previous data and determine the coupling energy of the dyad $E_{cpl}^{f-o}$ equal zero in both cases. This puts a serious problem for the explanation of the third stage of the above scheme forcing to suggest the origination of a peculiar IMI between $C_{60}$ and $O_2$ molecules in the excited state once absent in the ground state.

However, the IMI in the PD solutions is not limited by the fullerene-oxygen (*f-o*) interaction only. There are two other interactions, namely: fullerene-fullerene (*f-f*) and fullerene-solvent (*f-s*), among which the former is quite significant thus revealing itself as the fullerene dimerization [55] as well as a considerable amplification of the spectral properties of fullerene solutions [56]. The *f-s* interaction in the case of aqueous and benzene solutions can be ignored. Besides a significant strength, the *f-f* interaction possesses some peculiar features caused by the exclusive D-A ability of fullerenes [49]. A significant contribution of the D-A component into the total IMI results in a two-well shape of the potential energy term of a pair of fullerene molecules in the ground state, which is schematically shown in Fig.5a. According to the scheme, the pairwise interaction between the molecules in convenient solutions always leads to the formation of bi-molecular or more complex homoclusters of fullerenes in the vicinity of the $R^{00}$ minimum on the potential energy curve. The dimerization (as well as oligomerization) is a barrier reaction and does not occur spontaneously. Particular measures should be undertaken to come over the barrier and provide the molecule chemical coupling whilst the dimerization is energetically profitable. Photoexcitation is one of the most efficient tools. Therefore the PD solutions under ambient conditions should involve conglomerates of clustered $C_{60}$ molecules as shown schematically in Fig.5b, which is experimentally proven in many cases (see for example [56-58]).

UHF calculations determine the coupling energy of the pairwise *f-f* interaction for $C_{60}$ as $E_{cpl}^{f-f}$ =-0.52 *kcal/mol*. If remember that $E_{cpl}^{f-o}$ =0 in both singlet and triplet state, it becomes clear that oxygen molecules do not interact with either individual fullerene molecule or the molecule clusters so that the total energy of any dyad [$(C_{60})_n$-$O_2$] ( n= 1, 2, 3....) is just a sum of those related to the dyad components. It is always by 9.93 kcal/mol less in the triplet state due to the difference in the energy of the triplet and singlet oxygen (the UHF energy of $^3O_2$ and $^1O_2$ molecules constitutes -27.75 and -17.82 kcal/mol, respectively). Therefore, the ground state of the dyads is triplet.

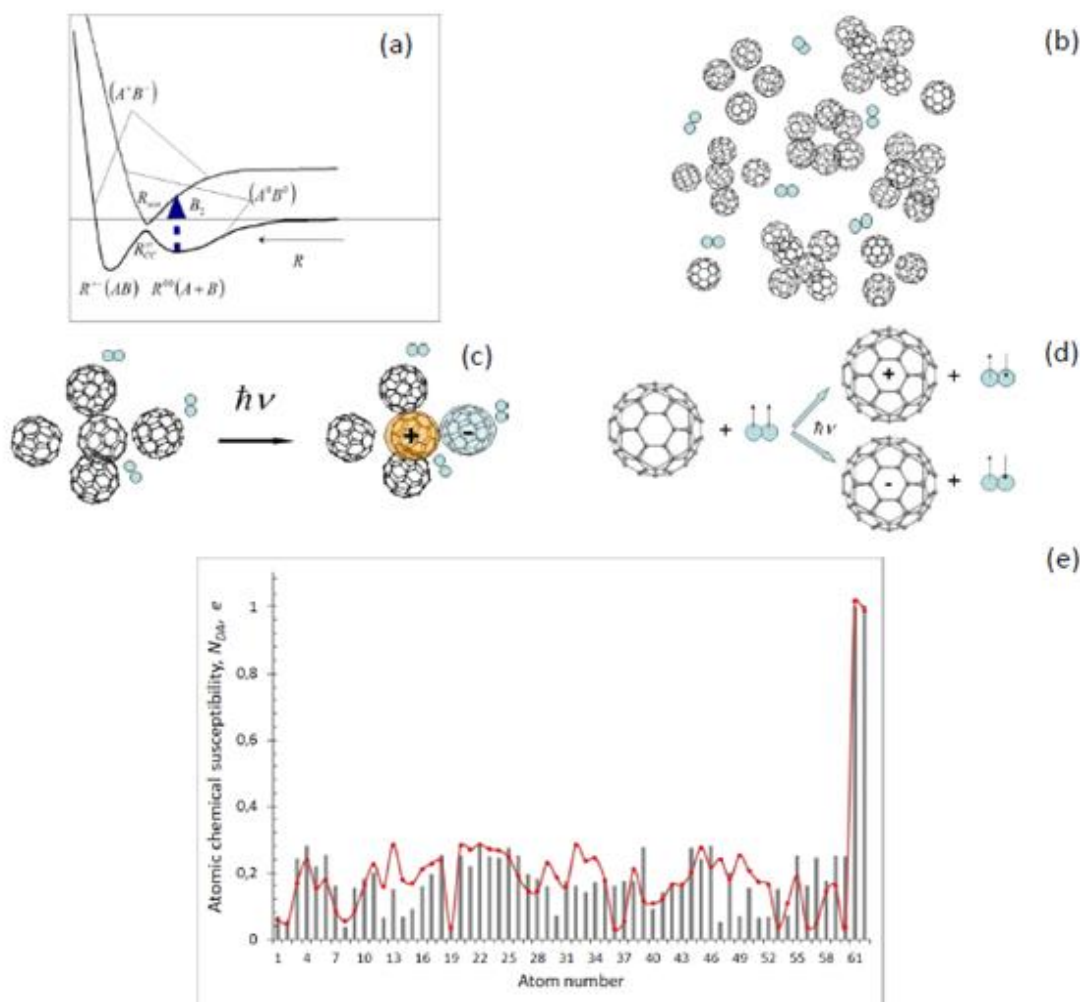

**Figure 5**. Photodynamic effect of the [$C_{60} + O_2$] complex. (a) Scheme of the IMI terms related to the interaction between two fullerene molecules [54]. $(A^0 B^0)$ and $(A^+ B^-)$ match branches of the terms related to the interaction between neutral molecules and their ions, respectively. $R^{+-}$ and $R^{00}$ mark minimum positions attributed to the formation of tightly bound dimer (AB) and weakly bound charge transfer complex ($A^0$+$B^0$), respectively. $R_{scn}$ indicates the point of avoidable intersection of terms $(A^0 B^0)$ and $(A^+ B^-)$. (b) Schematic presentation of fullerene clusterization in solution. (c) The formation of an ionic pair of $C_{60}$ under photoexcitation. The only ion pair is formed under photoexcitation of the $(C_{60})_n$ cluster of any content. (d) Schematic presentation of the spin-flip in oxygen molecule under photoexcitation. (e) Distribution of atomic chemical susceptibility over atoms of $^2[C_{60}^- + O_2]$ (histogram) and $^2[C_{60}^+ + O_2]$ (dotted curve) complexes.

Computations have shown [55, 56] that each pair of fullerene molecules as well as any fullerene cluster of more complex structure formed at the $R^{00}$ minimum is a charge transfer complex. Their absorption bands related to $B_2$ phototransitions in Fig.5a are located in the UV-visible region. The photoexcitation of either pair or cluster of fullerene molecules within this region produces a pair of molecular ions that quickly relax into the ground state of neutral molecule after the light is switched off. Calculations revealed [54] that, in contrast to neutral $C_{60}$, both molecular ions $C_{60}^-$ and $C_{60}^+$ actively interact with oxygen molecule producing coupling energy $E_{cpl}^{-f-o}$ and $E_{cpl}^{+f-o}$ of -10.03 and -10.05 *kcal/mol*, respectively, referring to $^3O_2$ molecule and -0.097 and -0.115 *kcal/mol* in regards to $^1O_2$. Therefore, the pristine oxygen molecule is quite strongly held in the vicinity of both molecular ions forming $[C_{60}+O_2]^-$ and $[C_{60}+O_2]^+$ complexes as schematically shown in Fig.5c. The complexes are of $^2[C_{60}^-+O_2]$ and $^2[C_{60}^++O_2]$ compositions of the doublet $SpM$. Both fullerene ions take the responsibility over the multiplicity, so that two odd electrons of the oxygen molecule 'lose their job' and do not more maintain the molecule triplet $SpM$ thus adding two effectively unpaired electrons to the $N_D$ pool of unpaired electrons of the whole complex as it is shown in Fig.5e. A dominant contribution of electrons located on oxygen atoms 61 and 62 is clearly seen thus revealing the most active sites of the complexes. It should be noted that these distributions are intimate characteristics of both complexes so that not oxygen itself but $^2[C_{60}^-+O_2]$ and $^2[C_{60}^++O_2]$ complexes as a whole provide the oxidative effect. The effect is lasted until the complexes exist and is practically immediately terminated when the complexes disappear when the light is switched off. The obtained results make it possible to suggest the PD mechanism schematically presented in Fig. 5d. As shown in the figure, changing $SpM$ from triplet to doublet one under photoexcitation due to passing from neutral molecule complex to those based on fullerene molecular ions results in a spin flip in the system of two odd electrons of the oxygen molecule. This approach allows attributing PD effect of fullerene solutions to a specific type of spin-chemical reactions. Particular D-A properties when the molecule can be both the donor and acceptor of electrons are evidently characteristic not only to fullerenes $C_{60}$ and $C_{70}$ as well as their derivatives, but to CNT [44] and nanosize graphene molecules [59] thus demonstrating them as additional characteristic feature of open-shell molecules. This opens a large new branch of a spin phochemistry of the species.

5. **Spin nature of the reaction final product**

So, equipped with high MCS and a large network of target atoms, the above-considered representatives of *sp²* nanocarbons enter into a chemical reaction. Leaving aside the DA interaction, we will consider below combination reactions in the language of $N_D$ and $N_{DA}$ emergents. In contrast to general chemistry, all the reactions of this kind involving *sp²* nanocarbons concern the molecules polyderivatization due to large network of targets. The latter are definitely discriminated by the relevant $N_{DA}$ value. The first steps in all the cases occur with the carbon atoms of the highest activity, which position is predetermined by UHF calculations: those are hexagon light blue atoms of $C_{60}$, open-end atoms of CNTs, and zigzag edge atoms of graphene molecules. A lot of information concerning open-end atoms of CNTs and edge atoms of graphene sheets can be found in literature. Thus, a convincing evidence that the edge atoms of graphene carry additional electronic (including spin) density was shown by unique experiments on scanning graphene ribbons with atomic resolution [60, 61] (see a detailed discussion of the experiments in [7]). In contrast, there is practically no information about particular hexagon packed atoms of fullerene. Nevertheless, considering the available information, we are inclined to conclude that convincing confirmation of the existence of special two rings in the fullerene structure was first discovered in 2008 and then confirmed in 2012 in atom-resolved STM [62] and AFM [63] studies. The joint results of the studies are presented in panel **I** in Fig.6. A scrupulous analysis of the STM image [62] showed that the molecule is attached to the Cu(111) substrate via contact area having a hexagon shape, therewith the area is duplicated above thus forming the imaged produced by scanning two hexagon circles of atom located one over the other. In the molecule structure there are only one pair of such hexagon configurations that coincide with those marked by light blue color in Fig. 1a. It is obvious that when the molecule interacts with an extended solid body, the coupling depends on the number of contacts and the interaction strength in each contact. Six light blue carbon atoms best satisfy both requirements, which explains the hexagon configuration of the molecule contact with the flat Cu(111) surface, as seen by STM. The experimental image of AFM scanning (**I**), in its turn, exhibits homogeneous distribution of the interaction force over the atoms which is expected if take into account the equality of $N_{DA}$ values for the atoms. A tight connection with exclusive atom-resolved AFM images (see a profound review [64]) and $N_{DA}$ distribution is discussed in details in Ref. [7].

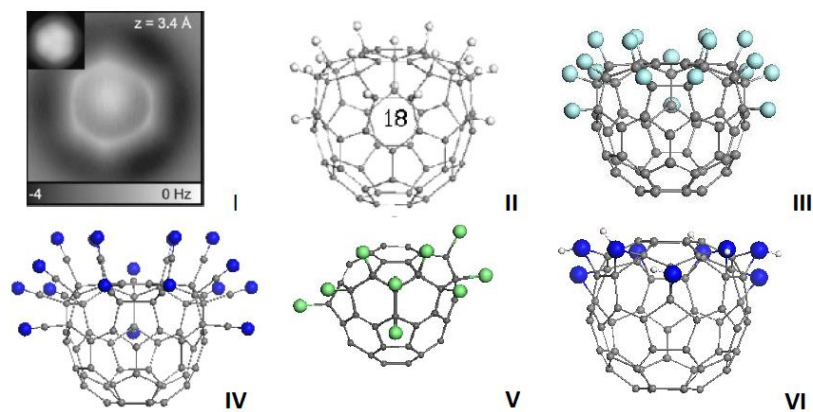

**Figure 6.** (**I**) STM image of $C_{60}$ on Cu(111) (left up corner) that reveals the orientation of $C_{60}$ [62] while the main image presents the constant-height AFM images of $C_{60}$ on Cu(111) obtained with a CO tip. Adapted from [63]. (**II – VI**) Computational spin chemistry of fullerene $C_{60}$. Monodentant polyderivatives $C_{60}H_{18}$ (**II**), $C_{60}F_{18}$ (**III**), $C_{60}(CN)_{18}$ (**IV**), and $C_{60}Cl_{18}$ (**V**) obtained at the 18$^{th}$ step of the successive polyderivatization following $N_{DA}$ algorithm; the same for bidentant poliderivative $C_{60}(NH)_9$ (**VI**) at the 9$^{th}$ step. See detailed discussion in [43].

When the first-order derivative is formed, the molecule responds to the action by reconstructions of its {$sp^2$C-C bonds} pool and redistribution of $N_{DA}$ markers. The newly appeared highest $N_{DA}$ mark determines the target carbon atom for the second step and so forth. Therefore, following the highest $N_{DA}$ as descriptor of the target atom at each reaction step determines the $N_{DA}$ algorithm of $sp^2$ carbon molecules polyderivatization. This algorithm is particularly important in the case of fullerenes and graphene molecules while the chemistry of CNTs is mainly the chemistry of their sidewalls [47].

Practical chemistry of $sp^2$ nanocarbons gives many examples confirming the $N_{DA}$ algorithm implementation. The most characteristic results are obtained for fullerene $C_{60}$ [43]. Let us illustrate some examples from the standpoint of the $N_{DA}$ algorithm. Exhibited in Fig. 6, polyderivatives II-V present a collection of one-dentant $C_{60}$ (X)$_{18}$ (X=H, F, CN, Cl) species. The first three molecules represent so-called crown structures of local symmetry $C_{3v}$. The structures were obtained at the 18$^{th}$ step of the successive addition of the addends to $C_{60}$ following $N_{DA}$ indication. All the above polyderivatives were synthesized in practice and their structures identical to shown in Fig. 6 were determined (see a detailed discussion of the issue in [43]). If the successive addition of addends in the case of $C_{60}H_{18}$ (**II**), $C_{60}F_{18}$ (**III**), and $C_{60}(CN)_{18}$ (**IV**) proceeds quite similarly, addition of chlorine behaves differently from the first step due to which the structure of $C_{60}Cl_{18}$ (**V**) turns out different. This result correlates with known problems concerning the fullerene chlorination [65]. $C_{60}(NH)_9$ (**VI**) molecule is a bi-dentant species and its structure at the 9$^{th}$ step of reaction is of $C_{3v}$ local symmetry thus showing a similar 'crown' configuration, which perfectly fit experimental data.

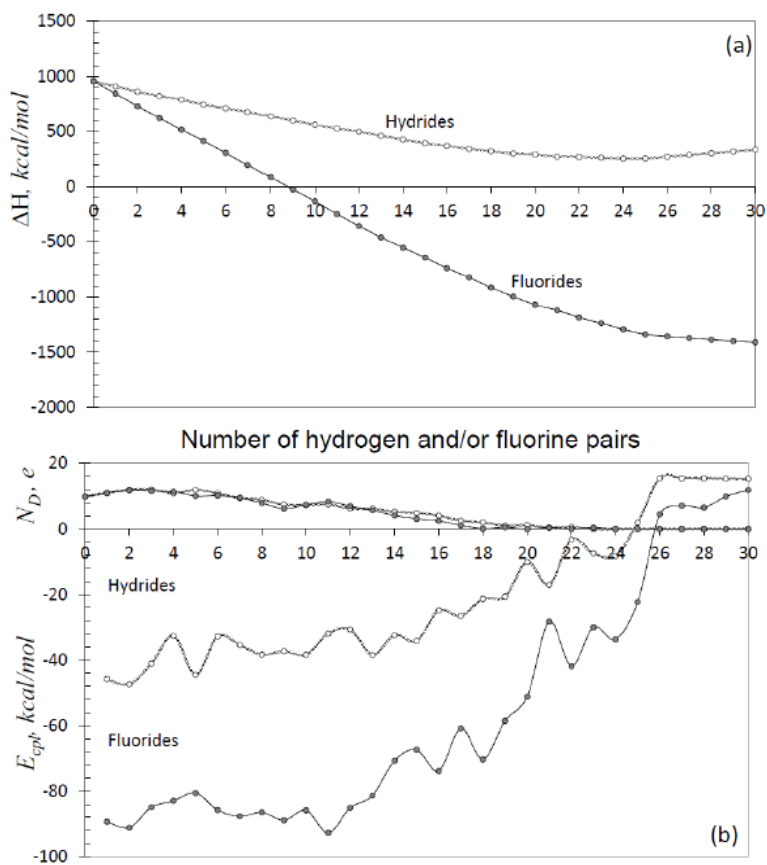

**Figure 7.** Evolution of the total energy $\Delta H$ (a), molecular chemical susceptibility $N_D$ and per-step coupling energy $E_{cpl}$ (b) at growing the number of atom pairs $k$ for $C_{60}$- hydrides (dotted curves with empty circles) and $C_{60}$- fluorides (solid curves with filled circles).

Another face of the $N_{DA}$ algorithm action can be seen from plottings in Fig. 7. Those present the evolution of the heat of formation, $\Delta H$, per-step coupling, $E_{cpl}$, and $N_D$ quantities in due course of hydrogenation and fluorination [66, 67]. As seen in the figure, the two reactions are characterized by significantly different energetic parameters with an obvious favoring to fluorination. Actually, the total energy $\Delta H$ of hydrides gradually decreases, which favors the polyhydrides formation, until $k$ reaches 15, after which the decreasing temp is slowed down approaching zero at $k$ = 25 and then the energy starts to slightly increase. Since for hydrogen $\Delta H_{mol} = 0$, the $E_{cpl}$ behavior directly exhibits the above changing in the total energy. The behavior of the total and coupling energies of fluorides is similar with the only difference concerning much bigger values of both $\Delta H$ and $E_{cpl}$ as well as the absence of the $\Delta H$ increasing at $k >25$. However, due to a significant negative value of $\Delta H_{mol}$ in this case, the coupling energy becomes positive at $k >25$.

Molecular chemical susceptibility $N_D$ is the other characteristic quantifier. As seen in Fig. 7b, $N_D(k)$ functions are practically identical for both families gradually decreasing at higher $k$ and approaching zero at $k\sim 20$-24. Therefore, decreasing $E_{cpl}$ by absolute value correlates with decreasing MCS $N_D$, or, by other words, with working out the pool of effectively unpaired electrons, which results in a considerable lowering of the reaction activity when $k$ changes from 18-20 to 25-26. According to both characteristics, the reaction is terminated at $k >25$. Important to note, that in spite of obvious obstacles, $k$-high products might be abandoned among the final products. This is due to accumulative character of the reaction until the next addition of the atom pair is still energetically favorable. This means that (1) the attachment of a next atom pair will not proceed at positive $E_{cpl}$ and (2) the accumulation time (and mass yield of the product) will greatly depend on the absolute value $E_{cpl}$: the less the value the longer time is needed. That is why more than four-times difference in the $E_{cpl}$ absolute values for hydrides in favor of fluorides at $k=18$ and its small absolute value result in the termination of hydrogenation process by $C_{60}H_{36}$ product while fluorination still continues and is completed by $C_{60}F_{48}$. The data presented in Fig. 7 are in full consent with experimental ones.

Presented in Fig. 8 briefly summarizes main features concerning polyderivatization of (5, 5)NGr molecule. The molecule is flat and consists of 66 carbon atoms. Chemical portrait of the molecule is given in the upper part of the figure as ACS $N_{DA}$ distribution over atoms. Following $N_{DA}$ algorithm, the formation of framed graphene molecules occurs at the first stage of the molecule chemical modification (see detailed discussion in [30, 31, 36]) after which the basal plane becomes the battlefield. Final products are produced in the course of topochemical reaction due to dependence on such important factors as 1) fixation or free standing of the molecule (membrane); 2) accessibility of either one or both sides of the membrane plane; 3) chemical addends in use. The role of these three factors can be seen in Fig. 8 where is exhibited a collection of (5, 5)NGr polyderivatives related to the maximum covering of the parent molecule by addends (in the molecules general inscription $C_{66}X_{2n2n}X_{km}$ (X=H, F, OH, and O), $X_{2n2n}$ and $X_{km}$ correspond to the termination of $n$ edge and $km$ basal-plane atoms, respectively). Thus, left column presents graphene hydrides obtained computationally in the course of successful hydrogenation of the parent molecule following $N_{DA}$ algorithm at different conditions. As seen, their structures differ drastically changing from regular graphane structure (a) to highly bent pancake (c) [68]. A canape-like structure of hydride obtained under one-side access of a fixed membrane completes a full list of the products. Experimentally regular graphane and amorphous canape-like structures of hydrogenated graphene membranes were observed under the required conditions [69]. When hydrogen is substituted by fluorine (see Fig. 8a), structure of polyfluorides changes drastically supporting, in general, (a)-(c) tendency of hydrides. However, none of regular graphene-like structures was obtained. A through experimental study of graphene fluorides named teflons [70] fully confirm this conclusion. The situation becomes still more cumbersome when hydrogen is substituted with hydroxyls (Fig. 8b) or a combination of oxygen atoms and hydroxyls (Fig. 8c). Last two cases present selected variants of numerous configurations related to graphene oxygenation [71].

Therefore, polyderivatization is extremely complicated and covers a great number of final products. As mentioned earlier, addition of each addend at every step is accompanied with immediate reconstruction of the {$sp^2$C-C bonds} pool. The pool is highly labile and readily responds to any perturbation. This $N_{DA}$- reconstructing distribution makes the polyderivatives production highly variable due to high sensitivity to any current situation, thus depriving such names as 'graphene hydride', 'graphene oxide', 'graphene fluoride' and so forth of an exact chemical formulae and allowing to speak about large classes of substances that are related to the above names. Therefore, the chemistry of graphene molecules is not a chemistry of a molecule, but a molecular class.

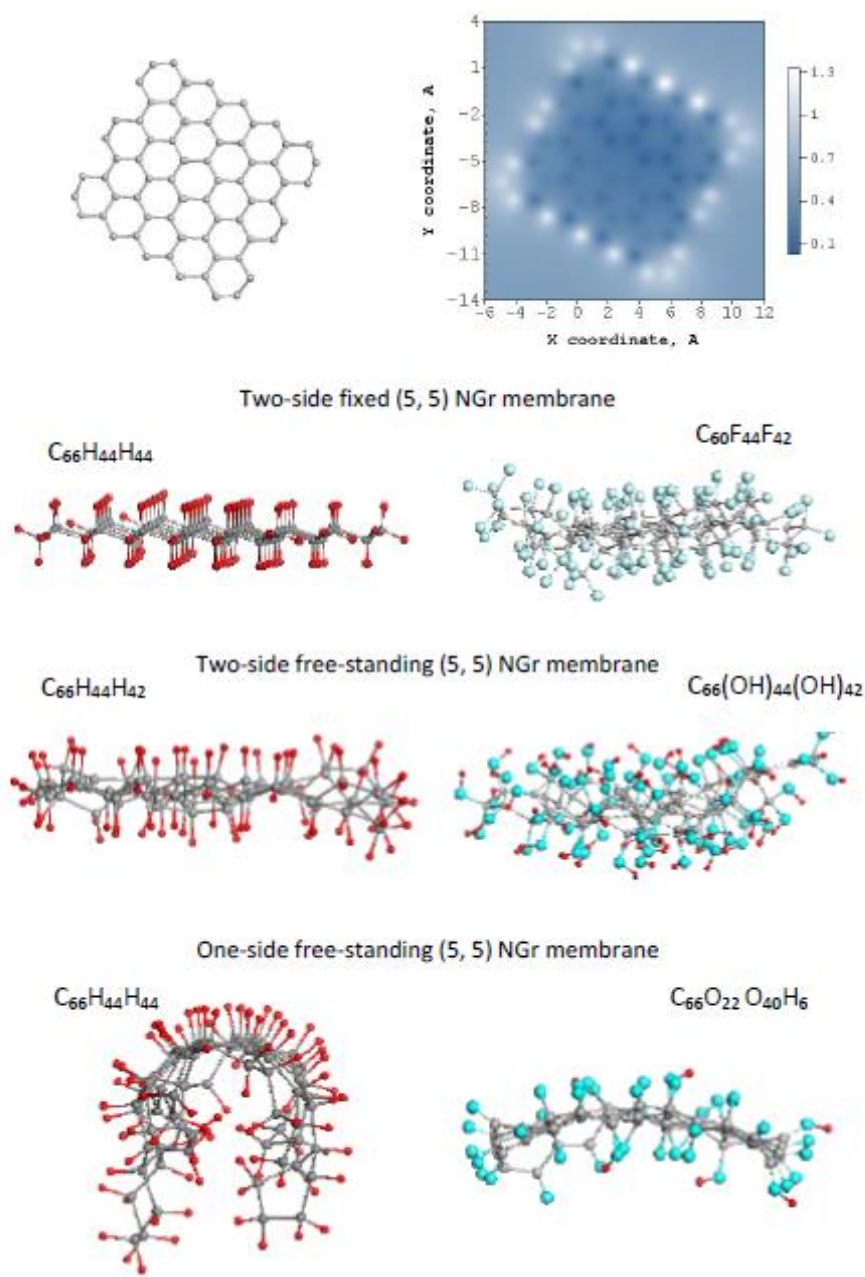

**Figure 8.** Top – Equilibrium structure of (5, 5)NGr molecule and the spatial map of the ACS $N_{DA}$ distribution over its atoms. (a)-(c). Equilibrium structures of complete graphene hydrides (left column), graphene fluoride, and two graphene oxides (right column, from top to bottom).

So far, we have dealt with products of the highest complete derivatization. The mass-produced 'graphene oxide' (GO) class is their best-known representative [72]. However, there are still innumerable products that are not completely derived. Framed graphene molecules are of such species. These molecules became a regular participant in the scientific agenda a decade ago when a new high-tech material, reduced graphene oxide (rGO), entered practical graphenics. The product is not synthesized from the bottom, but is produced from the top using GO as a parent species. Evidently, a large GO class gives life to a new rGO class, additionally subjected to variations due to variable conditions of the reduction procedures (see a profound review [73]). rGO molecules present fragments of the honeycomb structure framed by different atomic groups, depending on the reduction technique. The mass product present a powder of amorphous structure consisting of layered stacks of a few-nanometer thickness and from first nanometers to submicron dimension in the lateral direction. Interlayer distance is close to that one of graphite thus justifying a flat graphene-like structure of the layers (see as example, a neutron scattering study of different-origin rGOs [74]). Accordingly, in the literature you can often find a reference to rGO as graphene due to which it was suggested to cite it as technical graphene [75].

For a long time, from the beginning of the graphene era, there was an opinion that chemical derivatives of graphene, mainly GO, rGO and teflon, came into life together with graphene itself. The first breakdown was caused by the establishment that natural amorphous carbon, known as *shungite carbon*, is a natural deposit containing millions tons of rGO [76]. Then similar conclusions were made about the natural anthraxolite [77] and anthracite [39]. In time, it came to the industrial multi-tonnage carbon blacks as well [78, 79]. Actually, rGO of different kinds considered above are the member of the amorphous carbon family as well. Therefore, known more than one-and-half thousands of years, amorphous carbon is undergoing a rebirth and appears as agglomerative compositions of framed graphene molecules, thus becoming a special subject of modern nanotechnology. A recent extended study was devoted to a detailed investigation of the structure and chemical composition of a set of selected amorphous carbons of the highest carbonization [78]. For the first time, reasonable models of these graphene molecules, which are the basic structural units (BSUs) of the studied solids and which correspond to both their structure and chemical composition, were suggested. A set of selected models is shown in Fig. 9. The same (5, 5)NGr molecule, which is commensurate with experimentally determined BSUs of shungite carbon, anthraxolite, and carbon black 632, was chosen as parent structure in all the cases. The choice allows drawing the main attention to the framing areas of the models without distracting it to the structure of the carbon core. As seen in the figure, BSUs of amorphous carbons belong to the class of graphene oxyhydrides, differing by the hydrogen content and chemical composition of oxygen containing groups.

Revealing the molecular structure of amorphous carbon significantly changes our view of the world. We live on a land containing a huge amount of amorphous carbon, we breathe air with a high content of combustion products of organic substances, we eat food as polluted as air, we drink water extracted from underground sources or rivers suffering from carbon pollution. Isn't it time to ask ourselves if we are aware that in most of these life cases we are dealing with radicalized graphene molecules? The considered spin chemistry of the molecules gives a clear answer to this question.

## 6. Post-reaction existence of the products

Knowing from the discussed earlier that the parent (5, 5)NGr molecule has a high degree of radicalization, let us consider the radical status of the models shown in Fig. 9. UHF calculations reveal that all the models are radicals. Characterizing by $N_D$ value, the molecules form a following series of data: 27.4 *e* (shungite carbon), 22.2 *e* (anthraxolite), 29.7 *e* (carbon black 632) and 30.9 *e* (carbon black 624) [78]. The quantities are quite big evidencing strong radicalization. As seen in Fig.9, black balls of the $N_{DA}$ portraits in the circumference areas of the molecules clearly exhibit a fully inhibited chemical ability of both heteroatoms and carbon atoms to which the former are attached. Nevertheless, the remaining part of the molecules circumferences still remains highly active, particularly in the case of CB624, and easily accessible to not only gaseous reagents, but to bulky ones as well, once even aggravated with sterical constrains in the latter case.

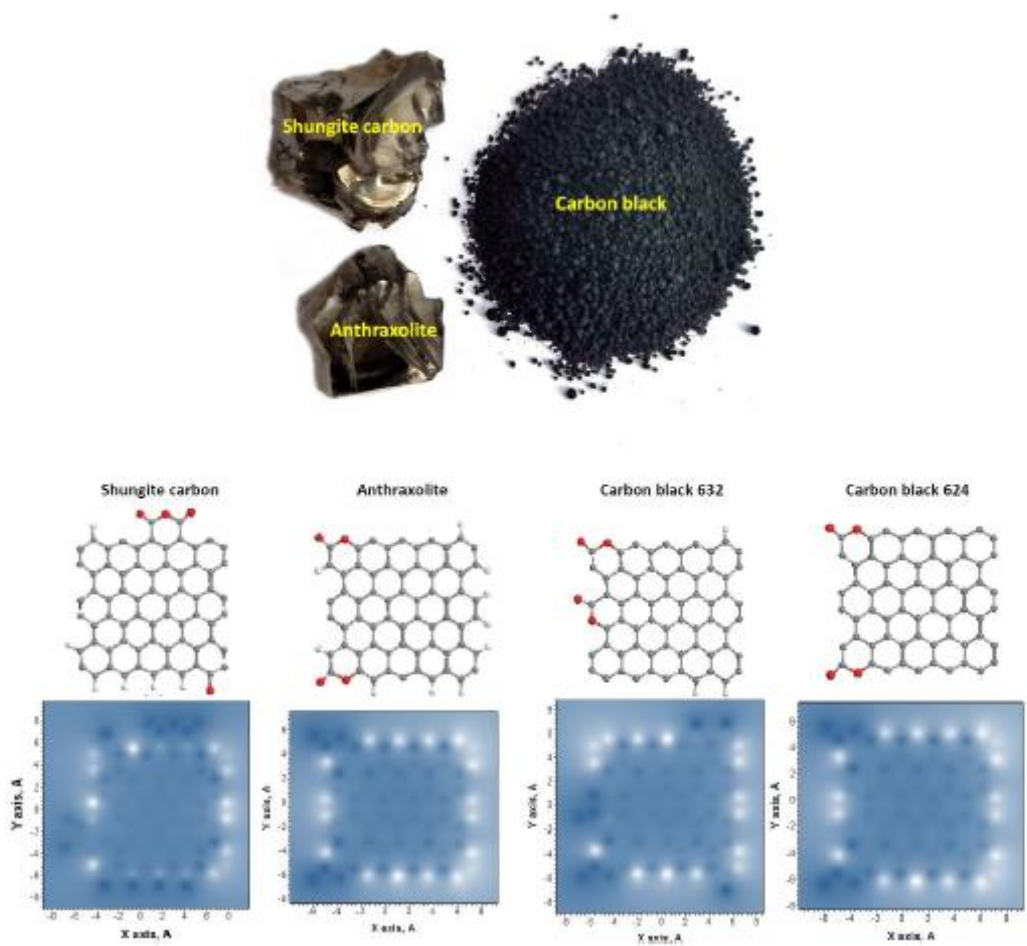

**Figure 9.** Appearance of natural amorphous carbon (top) and equilibrium structures of molecular BSU models (middle): *(a)* graphene oxihydrides $C_{68}O_4H_6$ attributed to "C=O" shungite carbon, (b) $C_{64}O_4H_{10}$ of "O=C-O-C'" antraxolite, (c) and (d) $C_{64}O_4H_3$ (c) and $C_{64}O_4$ (d) of "C-O-C" carbon blacks CB632 and CB624, respectively. (bottom) and $N_{DA}$ distribution over atoms of BSU model molecules (see text). UHF calculations. Adapted from [78].

Due to high radicalization of each suggested model and since the models, in general, are fully consistent with the discussed structural-compositional characteristics of the studied ACs [78], the authors made an astonishing conclusion that $sp^2$ amorphous carbons are conglomerates of stable radicals. The conclusion becomes still more impressive on the background of existing presentations that stable radicals, which are usually attributed to quite small molecules, are not ordinary part of the modern chemistry but are rare exclusion, the nature of which has still remained unclear (see reviews [39, 80-83] and references therein). On the other hand, it would seem that the huge amount of $sp^2$ amorphous carbons, both natural and engineered, as well as their exclusively broad involvement in various chemical and technical processes for more than a thousand of years could not disregard such an important characteristic of the substance if it existed. Nevertheless, in fact, by now there have not been communications joining radical terminology and huge massive of the available data concerning $sp^2$ amorphous carbons. At the same time, specific peculiarities, concerning heterogeneous catalysis [84-95] and electron-spin resonance [96-101], have been really observed and widely used. However, until now they have not been connected with specific radical properties of $sp^2$ carbons as well. Against this background, a study [102] that voiced the term "radical" in relation to $sp^2$ amorphics for the first time when attempting to explain carbonaceous soot inception and growth in term of resonance-stabilized hydrocarbonradical chain reactions, is quite revolutionary.

Restoration of BSUs radicals in their rights confronts us with new problems to be solved and the first one is to understand why $sp^2$ carbon molecular radicals are stable. The same should be attributed to fullerenes and CNTs, the radical essence of which has been convincingly established [43]. Since all open-shell molecules are radicals, the question of their stabilization has been raised repeatedly (see reviews [22, 23, 103]). However, the time-dependent stability of molecules was determined not directly but by a qualitative comparison of calculated thermodynamic factors. The procedure usually referred to molecules with terminated DBs, such as PAHs or *peri*-acenes. For these molecules it was established that spin-delocalized character of the molecule radicalization provided by the conjugation of $sp^2$ electrons over the total number of carbon atoms; nearly degenerated spin-triplet gap $E_{ST}$; incorporation of heteroatoms (O, N, S) inside benzenoid units or outside the latter favor the stabilization. All these factors are typical for graphene molecules, both bare and framed, such as, say, (5, 5)NGr and *peri*-antracene in Fig. 5 as well as semi-bare or/and semi-framed BSU molecules in Fig. 9. Only about the latter, we know exactly that they do live long [78].

Another problem of these molecules concerns real implementation of BSU structures from the available variety of possible models [78]. The models total energies $E_{gr}$ evidence a potential ability only while the nature of empirical observations lies in the chemical kinetics rather than the thermodynamic stability of the products. Reaction occurrence is governed by kinetic parameters that control the interrelation between reactants and products, forming free energy basins separated by barriers of different high. One can easily imagine how complicated the picture of the basins is when one of the reactant is multi-target. Until now, a possibility to quantitatively consider the relevant problems related to the formation of, say, shungite carbon BSUs, has seemed impossible. However, the appearance of a new method, called by its authors 'a multi class harmonic linear discriminant analysis' (MC-HLDA) [104], presenting metadynamics with discriminants as a tool for understanding chemistry, inspires some optimism that in the near future similar complex problems can be solved.

The idea of a particular role of kinetics in the case of multi-target chemical compounds with highly delocalized spin density puts time in the avant-garde row of particularly significant parameters. The chemistry of $sp^2$ nanocarbons requires the determination of the time of life of manufactured products. Despite the revolutionary contribution of $sp^2$ nanocarbons to modern chemistry, this area itself is still young, so the question of lifetime may seem premature. Nevertheless, evidence of the legitimacy of such an issue has already been obtained in synthetic chemistry. Thus, it was found that the lifetime of the simplest $sp^2$radicals, which are well-known PAHs, is different. Long-lived members of this series are naphthalene, anthracene, tetracene, pentacene, but the lifetime of higher PAHs drastically shortens when the number of cycles increases, so that hexacene has been still observed for a short time while the higher species cannot even be recorded as a result of chemical synthesis [23, 105, 106]. Synthesis of multinuclear hydrocarbons imitating graphene molecules by Müllen's team [107] also faces the problem of sustainability of the final products. The undoubted success of the synthesis is the existence of stable graphene materials, such as mass-produced GO, rGO, and graphene quantum dots based on it. Despite these materials have not been even two decades old and their temporal stability has not yet been truly investigated, the time-instability of GO has been rigidly fixed [74].

Besides synthetic laboratories, there is a unique natural laboratory, for which carbon is an absolute favorite. The laboratory operates in scale of time in billions (shungite carbon) and millions (anthraxolite, anthracite) of years. The time is large enough to expect that all chemical transformations, which could occurred with *sp²* carbons, already held. Consequently, compounds exist in size, shape and chemical composition, which no longer change. A careful analysis of these properties allows us to highlight a number of unique characteristics of natural amorphous carbons. 1) The average BSUs size of shungite carbon, anthraxolite and anthracite is of 1.5 nm nondependent on the local place of deposits [74, 78]. 2) The chemical composition of these mineral BSUs is similar [39]. 3) The BSUs preserve their radical character, which can be opened by changing the conditions of existence and/or storage of the substances. The latter lays the foundation of particular features of shungite carbon favoring its application in medical treatment [108], biology [109], non-linear optics [110], mechanics [111] and so on [112]. So far, anthracite and anthraxolite have not been known from this side while many exciting discoveries can be expected.

It becomes obvious as well that the properties of shungite carbon itself and in the vicinity to other minerals should be different due to a possible activation of radical properties of the former. Actually, any mineral nucleus is an atom cluster covered with DBs, thus having a radical nature, saturation of which when stacking with other nuclea promotes the mineral growth. Obviously, the termination of the bonds by another's addends terminates the growth [37]. Let's demonstrate the said on the example of the growth of a particular quartz in the presence of shungite carbon. $(SiO_4)_{48}$ cluster in Fig. 10 a presents a silica nucleus leading to the growth of $\alpha$-quartz. The cluster is composed in such a way that the consequent growth can occur from the basal top plane of the cluster that accommodates 9 silicon atoms with one DB each. The shungite carbon BSU molecule (Fig. 10b), which appears nearby, reacts with the cluster almost barrierless and is bound to it by two Si–C bonds, irrespective of the initial arrangement of the reactants (Figs. 10c and 10 d). The coupling energy constitutes -139.93and -124.35 kcal/mol, respectively. Evidently, the presence of the molecule prevents from joining two clusters for the quartz growth to be continued. When the molecule number is big, they together may form a 'patchwork fabric' to cover or envelope the silica nucleus. Assuming the nucleus spherical shape, geometry allows evaluate the sphere radius, when it is fully covered by the fabric consisting of ~1.5x1.5 nm sheets, which gives the value of 80-100 nm. Therefore, it is possible to expect the formation of a mineral matter consisting of peculiar 'chocolate balls' with a hard silica core inside, surrounded by schungite-carbon shell.

The problem under discussion did not arise from scratch. It was stimulated by a thorough physicochemical study of conditional shungite rocks with low carbon content [113]. Such a component is always present in shungite deposits. As occurred, the mineral matter actually consists of agglomerates of close-to-spherical nanoparticles presenting a crystalline inner core of perfect $\alpha$-quartz surrounded by shungite-carbon shell. The size of quartz core is of ~80 nm irrespectively of the deposits place. The shell can be easily separated from the core. The study of this carbon mass has revealed its full similarity to that one related to high-carbon shungite rocks. It is evident, that this first observing of a particular symbiosis of shungite carbon and silica should not be the only one. It can be assumed that the discovered phenomenon, which is self consistently explained from the standpoint of spin chemistry of nanocarbons, opens a new page in the history of geology, devoting it to spin geochemistry.

## 7. Conclusion

The paper presents a self-consistent overview of *sp²* nanocarbons including fullerenes, carbon nanotubes, and graphene molecules from the viewpoint of an experienced long-working user of computational quantum chemistry. The consideration is suggested to be performed basing on concepts of the UHF approximation in terms of its emergents that are originated by the spin symmetry breaking. Among the others, UHF emergent such as $SpD_{tot}$ and $SpD_A$ as well as $N_D$ and $N_{DA}$ ($SpD_{tot}$ and $N_D$ describe total spin density and total number of effectively unpaired electrons while subscript *A* matches these values related to atom A) were used.

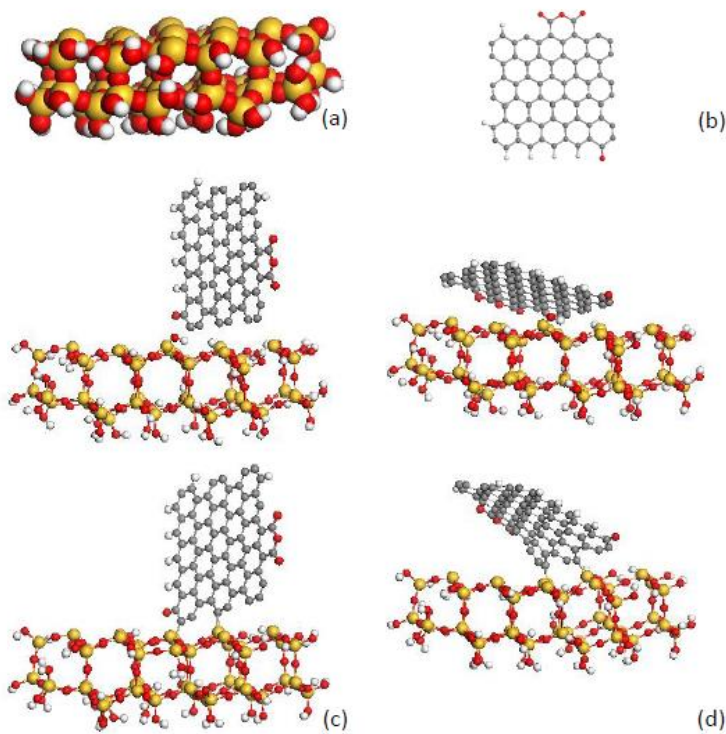

**Figure 10.** Equilibrium structures of $(SiO_4)_{48}$ (a) and $C_{66}H_6O_4$ graphene molecule (b) imitating α-quartz nucleus and BSU of shungite carbon, respectively. (c)-(d) Equilibrium structures of silica - shungite carbon composites (bottom) at two starting positions (top) of the components. UHF calculations.

High efficiency of the approach allowed performing a large set of computational experiments that provide getting a large number of necessary data for a completed vision of spin chemistry of *sp²* nanocarbons to be constructed. In light of these parameters, the above molecules appear as open-shell electronic systems with broken spin symmetry, radicalized multi-target objects of ongoing reactions, characterized by both spin density and chemical activity delocalization over molecules' atoms. This feature determines the existence of strict $N_D$ and $N_{DA}$ algorithms that control the course and final products of chemical reactions involving the species. A number of examples exhibiting main traits of computational spin chemistry of fullerene $C_{60}$, a fragment of (4, 4) SWCNT, and a family of graphene molecules on the basis of a parent (5, 5)NGr one is presented. A perfect fitting of calculated and experimental data is discussed. Particular attention is given to graphene molecules presenting basic structure units of synthetic and natural amorphous carbons. The stable-radical character of the latter allows rising question about spin geochemistry due to the ubiquitous presence of amorphous carbon in the Nature.

**Acknowledgements**

The author is greatly thankful to Ye. Golubev, N. Rozhkova, I. Natkaniec for fruitful discussions. The publication has been prepared with the support of the "RUDN University Program 5-100".

**References**

1. J.A. Pople and R.K. Nesbet, *J. Chem. Phys.*, **22**, 571 (1954).
2. P-O. Löwdin, Phys. Rev. **97**, 1509 (1955).
3. P-O. Löwdin, *Adv. Chem. Phys.,* **2**, 209 (1958).
4. S.S. Ray, S. Manna, A. Ghosh, R. K. Chaudhuri, and S.Chattopadhyay, *Int. J. Quant. Chem.* e25776 (2018).
5. K. Takatsuka, T. Fueno, and K. Yamaguchi, *Theor. Chim. Acta*, **48**, 175 (1978).
6. V.N. Staroverov and E.R.Davidson, *Chem. Phys. Lett.*, **330**, 161 (2000).
7. E.F. Sheka, N.A. Popova, and V.A. Popova, *Physics; Uspekhi* **61,** 645 (2018).
8. P.W. Anderson, *Science* **177** 393 (1972).
9. R.B. Laughlin, *Rev. Mod. Phys.* **71**, 863 (1999).
10. R.B. Laughlin and D. Pines, *Proc. Natl. Acad. Sci. USA* **97**, 28 (2000).
11. C. Yannouleas and U. Landman, *Rep. Prog. Phys*. **70**, 2067 (2007).
12. *Emergent Phenomena in Correlated Matter,* Autumn School, Jülich, 23±27 September 2013, edited by E. Pavarini, E.Koch, and U. Schollwöck (Forschungs-zentrum Jülich and the German Research School for Simulation Sciences, 2013).
13. M.V. Putz, O. Ori, M.V, Diudea, B. Zefler, and R. Pop, in *Distance, Symmetry, and Topology in Carbon Nanomaterials* (Carbon Materials: Chemistry and Physics, Vol. 9), edited by A.R. Ashrafi and M.V.Diudea) (Springer, 2016) 345.
14. Y. Kitagawa, T. Saito, Y. Nakanishi, Y. Kataoka, T. Matsui, T. Kawakami, M. Okumura, and K. Yamaguchi, J. Phys. Chem. A, **113**, 15041 (2009).
15. Y. Cui, I.W. Bulik, C.A. Jiménez-Hoyos, T.M. Henderson, and G.E. Scuseria, J. Chem. Phys. **139**, 154107 (2013).
16. Y. Yang, E.R. Davidson, and W. Yang, Proc. Natl. Acad. Sci. **113**, E5098 (2016).
17. J. Hachmann, J.J. Dorando, M. Aviles, and G.K.-L. Chan, J. Chem. Phys. **127**, 134309 (2007).
18. D. Casanova and M. Head-Gordon, Phys. Chem. Chem. Phys. **11**, 9779 (2009).
19. C.R. Jacob and M. Reiher, Int. J. Quantum Chem. **112**, 3661 (2012).
20. I.G.Kaplan, Int. J. Quant. Chem. **107**, 2595 (2007).
21. I.G.Kaplan, Mol. Phys. **116**, 658 (2018).
22. J. Shee, E. J. Arthur, S. Zhang, D. R. Reichman, and R. A. Friesner, J. Chem. Theory Comput., DOI: 10.1021/acs.jctc.9b00534 (2019).
23. T. Y. Gopalakrishna, W. Zeng, X. Lu and J. Wu, Chem. Comm. 54, 2186 (2018).
24. A.C.Ferrari and 63 more authors, Nanoscale, **7**, 4598 (2015).
25. K. S. Novoselov, V. I. Fal'ko, L. Colombo, P. R. Gellert, M. G. Schwab, and K. Kim, Nature **490**, 192 (2012).
26. A.P. Kauling, A. T. Seefeldt, D. P. Pisoni, R. C. Pradeep, R. Bentini, R. V. B. Oliveira, K. S. Novoselov, and A. H. Castro Neto, Adv. Mater. 1803784 (2018).
27. M. E. Sandoval-Salinas, A. Carreras, and D. Casanova, Phys. Chem. Chem. Phys., **21**, 9069 (2019).
28. E.F. Sheka and L.A. Chernozatonskii, Int. Journ. Quant. Chem. **110**, 1938 (2010).
29. E. F. Sheka, in *Nanoscience and Nanotechnologies*, edited by V. Kharkin, C. Bai, O. O. Awadelkarim, S. Kapitaza (Eolss Publishers, 2011).
30. E.F.Sheka, Int. J. Quant. Chem. **112**, 3076 (2012).
31. E.F.Sheka in *Advances in Quantum Methods and Applications in Chemistry, Physics, and Biology*, edited by M. Hotokka, E. Brändas, and J, Maruani (Spinger, 2013) 249.
32. E.F.Sheka, Int. J. Quantum Chem. **114**, 1079 (2014).

33. E. F. Sheka in *Quantum Systems in Physics, Chemistry, and Biology: Advances in Concepts and Applications,* edited by A. Tadjer, R. Pavlov, J. Maruani, E. Brändas, and G. Delgado-Barrio (Springer, 2017) 39.
34. E.F.Sheka, M.F.Budyka, and N.A.Popova, Rev. Adv. Mat. Sci. **51**, 35 (2017).
35. E.F.Sheka, Rev. Adv. Mat. Sci. **53**, 1 (2018).
36. E.F.Sheka. *Spin Chemical Physics of Graphene* (Pan Stanford, 2018).
37. R. Hoffmann, Ang. Chem. Int. Ed. **52**, 93 (2013).
38. E.F. Sheka, B.S. Razbirin, and D.K. Nelson, J. Phys. Chem. A, **115**, 3480 (2011).
39. E.F. Sheka, preprint arXiv:1909.04502 (2019).
40. E.F. Sheka, Adv. Quant. Chem. **70**, 111 (2015).
41. V.A. Zayets, *CLUSTER-Z1: Quantum-Chemical Software for Calculations in the s,p-Basis* (Inst. Surf. Chem. Nat. Ac. Sci. of Ukraine, 1990) (in Russian).
42. H. Zabrodsky, S. Peleg, and D.J. Avnir, Am. Chem. Soc. **114**, 7843 (1992).
43. E.F.Sheka, *Fullerenes: Nanochemistry, Nanomagnetism, Nanomedicine, Nanophotonics* (T&F CRC Press, 2011).
44. E.F. Sheka and L.A. Chernozatonsky, Int. Journ. Quant. Chem. **110**, 1466 (2010).
45. E.F. Sheka and E.V. Orlenko, Full. Nanot. Carb. Nanostr. **25**, 289 (2017).
46. E.F.Sheka, Journ. Str. Chem. **47**, 593 (2006).
47. F. Hauke and A. Hirsch in *Carbon Nanotubes and Related Structures Synthesis, Characterization, Functionalization, and Applications*, edited by D. M. Guldi and N. Martín (Wiley-VCH, 2010) 135.
48. E.F. Sheka, in *Topological Modelling of Nanostructures and Extended Systems*. Carbon Materials: Chemistry and Physics, Vol. 7, edited by A.R. Ashrafi, F. Cataldo, A. Iranmanesh, and O. Ori (Springer, 2013) 137.
49. E.F. Sheka, Int. J. Quant. Chem. **100**, 388 (2004).
50. F. Kasermann and C. Kempf, Rev. Med. Virol. **8**, 143 (1998).
51. L.B. Piotrovski in *Carbon Nanotechnology*, edited by. L. Dai (Elsevier, 2006) 235.
52. T. Da Ros in *Medicinal Chemistry and Pharmacological Potential of Fullerenes and Carbon Nanotubes,* edited by F. Cataldo and T. Da Ros (Springer, 2008) 1.
53. P. Mroz, G.P. Tegos, H. Gali, T. Wharton, T. Sarna, and M.R. Hamblin, in *Medicinal Chemistry and Pharmacological Potential of Fullerenes and Carbon Nanotubes,* edited by F. Cataldo and T. Da Ros (Springer, 2008) 79.
54. E.F. Sheka, Nanosci. Nanothech. Let. **3**, 28 (2011).
55. E.F. Sheka, Chem. Phys. Lett. **438**, 119 (2007).
56. E.F. Sheka, B.S. Razbirin, A.N. Starukhin, D.K. Nelson, M.Yu. Degunov, P.A. Troshin, and R.N. Lyubovskaya, J. Nanophot. SPIE **3**, 033501 (2007).
57. S. Nath, H. Pal, D.K. Palit, A.V. Sapre, and J.P. Mitta, J. Phys. Chem. B **102**, 10158 (1998).
58. S. Samal and K.E. Geckeler, J. Chem. Soc. Chem. Commun. 2224 (2001).
59. B.S. Razbirin, N.N Rozhkova, E.F. Sheka, D.K. Nelson, and A.N. Starukhin, J. Exp. Theor. Phys. **118**, 735 (2014).
60. J. van der Lit, M. P. Boneschanscher, D. Vanmaekelbergh, M. Ijäs, A. Uppstu, M. Ervasti, A. Harju, P Liljeroth, and I. Swart, Nat. Commun. **4**, 2023 (2013).
61. J.H. Warner, Y.C. Lin, K. He, M. Koshino, and K. Suenaga, Nano Lett. 14, 6155 (2014).
62. K.J. A. Larsson, S. D. Elliott, J. C. Greer, J. Repp, G. Meyer, and R. Allenspach, Phys. Rev. B **77**, 115434 (2008).
63. L. Gross, F. Mohn, N. Moll, B. Schuler, A. Criado, E. Guiti, D. Peça, A. Gourdon, and G. Meyer, Science **337**, 1326 (2012).
64. L. Gross , B. Schuler, N. Pavliček, S.Fatayer, Z. Majzik, N. Moll, D. Peña, and G. Meyer, Angew. Chem. Int. Ed. **57**, 2 (2018).

65. P.A. Troshin O.A. Troshina, R.N. Lyubovskaya, and V.F. Razumov, *Functional Derivatives of Fullerenes. Synthesis and Applications to Organic Electronics and Biomedicine* (in Russian) (Ivanovo State University, 2009).
66. E.F.Sheka, J. Mol.Mod. **17**, 1973 (2011).
67. E.F.Sheka, J. Exp. Theor. Phys. **111**, 395 (2010).
68. E.F.Sheka and N.A.Popova, J. Mol.Mod. **18**, 3751 (2012).
69. D.C. Elias, R.R. Nair, T.M.G. Mohiuddin, S.V. Morozov, P. Blake, M.P. Halsall, A.C. Ferrari, D.W. Boukhvalov, M.I. Katsnelson, A.K. Geim, and K.S. Novoselov, Science **323**, 610 (2009).
70. R.R. Nair, W. Ren, R. Jalil, I. Riaz, V.G. Kravets, L. Britnell, P. Blake, F. Schedin, A.S. Mayorov, S. Yuan, M.I. Katsnelson, H.M. Cheng, W. Strupinski, L.G. Bulusheva, A.V. Okotrub, I.V. Grigorieva, A.N. Grigorenko, K.S. Novoselov, and A.K. Geim, Small **6**, 2877 (2010).
71. E.F. Sheka and N.A. Popova, Phys. Chem. Chem. Phys. **15**, 13304 (2013).
72. J. Zhao, L. Liu, and F. Li, *Graphene Oxide: Physics and Applications* (Springer, 2015).
73. A.Y.C. Eng, C.K. Chua, and M. Pumera, Nanoscale, **7**, 20256 (2015).
74. Natkaniec, E.F. Sheka, K. Drużbicki, K. Hołderna-Natkaniec, S.P. Gubin, E.Yu. Buslaeva, and S.V. Tkachev, J. Phys. Chem. C **119**, 18650 (2015).
75. E.F. Sheka, and E.A. Golubev, Tech. Phys. **61**, 1032 (2016).
76. E.F. Sheka and N.N. Rozhkova, Int. J. Smart. Nanomat. **5**, 1 (2014).
77. Ye. A. Golubev, S. I. Isaenko, A. S. Prikhodko, N. I. Borgardt, and E. I. Suvorova, Eur. J. Mineral. **28**, 545 (2016).
78. Ye.A. Golubev, N. N. Rozhkova, E. N. Kabachkov, Y. M. Shul'ga, K. Natkaniec-Hołderna, I. Natkaniec, I. V. Antonets, B. A. Makeev, N. A. Popova, V. A. Popova, E. F. Sheka, J. Non-Cryst. Sol. **524,** 119608 (2019).
79. E.F. Sheka, K. Hołderna-Natkaniec, I. Natkaniec, J. X. Krawczyk,Ye. A. Golubev, N. N. Rozhkova, V.V. Kim, N. A. Popova, and V. A. Popova, J. Phys. Chem. C **123**, 15841 (2019).
80. H. Fisher, Chem. Rev. **101**, 3581 (2001).
81. *Stable Radicals: Fundamentals and Applied Aspects of Odd-Electron Compounds,* edited by R.G.Hicks (Wiley, 2010).
82. S. Müllegger, M. Rashidi, M. Fattinger, and R. Koch, J. Phys. Chem. C **116**, 22587 (2012).
83. X. Qi, L. Zhu, R. Bai, and Y. Lan, Sci. Rep. **7**, 43579 (2017).
84. L.R. Radovich in *Chemistry and Physics of Carbon*, vol. 27, edited by L.R. Radovich (Marcel Dekker, 2001) 131.
85. *Carbon Materials for Catalysis*, edited by P. Serp and J.L. Figueiredo (Wiley, 2009).
86. *Nanostructured Carbon Materials for Catalysis,* edited by P. Serp and B. Machado (Royal Society of Chemistry, 2015).
87. L.R. Radovich, in *Carbon Materials for Catalysis*, edited by P. Serp and J.L. Figueiredo (Wiley, 2009) 1.
88. R.C. Bansal and J.-B. Donet in *Carbon Black. Science and Technology*, edited by J.-B. Donet and R.C. Bansal (Marcel Dekker, 1993) 175.
89. *Metal-Free Functionalized Carbons in Catalysis: Synthesis, Characterization and Applications*, edited by A. Villa and N. Dimitratos (Royal Society of Chemistry, 2018).
90. T.J. Bandosz, in *Carbon Materials for Catalysis*, edited by P. Serp and J.L. Figueiredo (Wiley, 2009) 45.
91. R.I. Bel'skaya, in Shungites – New Carbon Raw Materials (Karelian Sci. Center, 1984) (in Russian).
92. E.N. Grigoyeva and N.N. Rozhkova, Russ. J. Appl. Chem. 73, 600 (2000) (in Russian).

93. N.N. Rozhkova, in *Perspectives of Fullerene Nanotechnology*, edited by E.Osawa (Springer, 2002) 237.
94. N.N. Rozhkova, L.E. Gorlenko, G.I. Yemel'yanova, A. Jankowska, M.V. Korobov, V.V. Lunin, and E. Osawa, Pure Appl. Chem. **81**, 2092 (2009).
95. F. Hu, M. Patel, F. Luo, C. Flach, R. Mendelsohn, E. Garfunkel, H. He and M. Szostak, J. Am. Chem. Soc. **137**, 14473 (2015).
96. C. Hu and L. Dai, Ang. Chem. Int. Ed. **55**, 11736 (2016).
97. P. Rapta, A. Bartl, A. Gromov, A. Stasoeko, and L. Dunsch, Chem. Phys. Chem. 3, 351 (2002).
98. M. Kosaka, T.W. Ebbesen, H. Hiura, and K. Tanigaki. Chem. Phys. Lett. 233, 47 (1995).
99. M.V. Trenikhin, O.V. Ivashchenko, V.S. Eliseev, B.P. Tolochko, A.B. Arbuzova, I.V. Muromtsev, Yu. G. Kryazhev, V.A. Drozdov, E. Sazhina, and V.A. Likholobov, Full. Nanot. Carb. Nanostr. **23**, 801 (2015).
100. S.V. Krasnovyd, A.A. Konchits, B.D. Shanina, M.Ya. Valakh, I.B. Yanchuk, V.O. Yukhymchuk, A.V. Yefanov, and M.A. Skoryk, Nanoscale Res. Lett. 10:78 (2015).
101. V.I. Silaev, V.O. Ilchenko, V.P. Lyutoev, V.N. Filippov, Ye. A. Golubev, and O.V. Kovaleva, in *Problems of Geology and Mineralogy*, edited by N.P. Yushkin (Geoprint, 2006) (in Russian) 283.
102. K.O. Johansson, M.P. Head-Gordon, P.E. Schrader, K.R. Wilson, and H.A. Michelsen, Science **361**, 997 (2018).
103. S. Gronert, J.R. Keeffe, and R.A. More O'Ferrall*,* in *Contemporary Carbene Chemistry*, edited by R.A. Moss and M.P.Doyle (Wiley, 2014) 3.
104. G.M. Piccini, D. Mendels, and M. Parrinello, J. Chem. Theory Comput. **14**, 5040 (2018).
105. H.F. Bettinger, Pure Appl Chem. **82**, 905 (2010).
106. F. Plasser, H. Pašalić, M.H. Gerzabek, F. Libisch, R. Reiter, J. Burgdörfer, T. Müller, R. Shepard and H. Lischka, Ang Chem, Int. Ed. **52**, 2581 (2013).
107. X.-Y. Wang, A. Narita, and K. Müllen, Nat. Rev. Chem. **2**, 0100 (2017).
108. V.A. Krutous, *Medicinal Properties of Shungite* (Karelia-shungite, 2016) (in Russian).
109. A.S. Goryunov, A.G. Borisova, and S.P. Rozhkov *Proc. Karelian Res. Center RAS, Exp. Biol. Ser.* **2**, 154 (2012).
110. N.V. Kamanina, S.V. Serov, N.A. Shurpo, N.N. Rozhkova, Tech. Phys. Lett. 37, 949 (2011).
111. N.V. Usol'tseva, M.V. Smirnova, A.V. Kazak, A.I. Smirnova, N.V. Bumbina, S.O. Il'in, and N.N. Rozhkova, J. Frict. Wear **36**, 380 (2015).
112. S. Politi, R. Carcione, E. Tamburri, R. Matassa, T. Lavecchia, M. Angjellari, and M. L. Terranova, Sci. Rep. **8**, 17045 (2018).
113. R.M. Sadovnichii, S.S. Rozhkov, and N.N. Rozhkova, Smart Nanocomp. 7, 111 (2016).